\UseRawInputEncoding
\documentclass[prd,12pt,onecolumn,superscriptaddress,nofootinbib]{revtex4-1}
\usepackage{mathtools}
\usepackage{amssymb}
\usepackage{amsfonts}
\usepackage[footnotesize]{caption}
\usepackage[font=scriptsize]{subcaption}
\usepackage{color}
\usepackage{braket}
\usepackage{hyperref}
\hypersetup{colorlinks,
citecolor=blue
}
\usepackage{url}
\usepackage{multirow}
\usepackage{relsize}
\usepackage{fullpage}
\usepackage{makecell}
\usepackage{fullpage}

\newcommand{\ba}{\begin{array}}
\newcommand{\ea}{\end{array}}
\newcommand{\be}{\begin{equation}}
\newcommand{\ee}{\end{equation}}
\newcommand{\beqa}{\begin{eqnarray}}
\newcommand{\eeqa}{\end{eqnarray}}
\def\321{$SU(3)\times SU(2)\times U(1)$}

\begin{document}

\title{\boldmath \bf New physics at nuSTORM}

\author{Kaustav Chakraborty}
\email[Email Address: ]{kaustav@prl.res.in}
\affiliation{Theoretical Physics Division, Physical Research Laboratory, Ahmedabad - 380009, India}
\affiliation{Discipline of Physics, Indian Institute of Technology, Gandhinagar - 382355, India}

\author{Srubabati Goswami}
\email[Email Address: ]{sruba@prl.res.in}
\affiliation{Theoretical Physics Division,
Physical Research Laboratory, Ahmedabad - 380009, India}

\author{Kenneth Long} 
\email[Email Address: ]{k.long@imperial.ac.uk}
\affiliation{Department of Physics, Imperial College London, Exhibition Road, SW7 2AZ, UK}
\affiliation{STFC Rutherford Appleton Laboratory,Harwell Campus, Didcot, OX11 0QX, UK}
%............

\begin{abstract}
{In this work we investigate the usefulness of nuSTORM as a probe of two new-physics scenarios which are sterile neutrinos and non-unitarity of the neutrino mixing matrix. For the sterile neutrino we show the importance of the neutral current events when combined with the charged current events to constrain the effective mixing angle, $\theta_{\mu \mu}$, and the sterile mixing angles $\theta_{14}$ and $\theta_{24}$.
We also study the role nuSTORM will play in the study of neutrino oscillation physics if the three generation neutrino mixing matrix is non-unitary. In this context we elucidate the role of nuSTORM, considering both charged current and neutral current events, in constraining the various non-unitarity parameters such as $\alpha_{11}$, $|\alpha_{21}|$ and $\alpha_{22}$.}  
\end{abstract}

\maketitle

\section{Introduction} 

Neutrino oscillation experiments have conclusively established the 
paradigm of the three-flavour neutrino oscillations and oscillation parameters 
are being determined with increasing precision. 
The three parameters that are yet to be determined are the mass hierarchy, the octant of the 
atmospheric mixing angle, $\theta_{23}$, and the leptonic CP phase, 
$\delta_{CP}$. There are some indications of the value of these parameters from the 
current data. Future planned/proposed high statistics experiments 
are expected to clinch these issues. 
With the determination of the three-neutrino mixing parameters already on the horizon, 
efforts have been made to explore new physics beyond the Standard Model 
in these experiments. 
New physics scenarios that have garnered considerable interest in the community 
include light sterile neutrinos, non-unitarity of the neutrino mixing matrix, 
non-standard interactions of the neutrinos etc. 

The existence of light sterile neutrinos was postulated to 
explain the LSND results \cite{Athanassopoulos:1996jb, Aguilar:2001ty}. LSND reported signals of 
$\nu_\mu-\nu_e$ oscillations with mass-squared difference of the order of eV$^2$. 
This was supported by MiniBooNE \cite{Aguilar-Arevalo:2013pmq, Aguilar-Arevalo:2018gpe} and also by the gallium and reactor anomalies \cite{Acero:2007su, Mueller:2011nm, Mention:2011rk}. 
In order to accommodate the $\rm{eV}^2$ oscillation scale the simplest possibility is to add a sterile neutrino to the Standard Model. There are two possible ways this can be done.— (i) The 2+2 scenario in which the oscillation to sterile neutrino constitute the dominant solution either to solar or atmospheric neutrino anomaly and disfavoured from current data \cite{Maltoni:2002ni}. 
(ii) The 3+1 or 1+3 picture in which the sterile neutrino is separated by an $\rm{eV}^2 $ mass difference  from the 3 active states \cite{Goswami:1995yq}. 3+1 (1+3)  corresponds to the 3 active states to be lighter (heavier).  Cosmological constraints on  sum of neutrino masses pose a serious challenge in accommodating an eV scale sterile neutrino scenario. To address these, secret neutrino interactions \cite{Chu:2018gxk} or lower reheating temperature \cite{Gelmini:2004ah, Yaguna:2007wi, deSalas:2015glj} are proposed. The 1+3 picture is more disfavoured from cosmology since there are three heavier states. 
%The 3+1 picture  can provide an acceptable fit  to the data \cite{Gariazzo:2017fdh, Dentler:2018sju} albeit the tension between disappearance and appearance data. 
The 3+1 picture  can provide an acceptable fit  to the data \cite{Gariazzo:2017fdh, Dentler:2018sju} albeit the tension between disappearance and appearance data. This tension is driven mainly by $\nu_\mu$ disappearance data and the LSND appearance data \cite{Maltoni:2002xd, Diaz:2019fwt}
%. This tension is driven by LSND data 
while the contribution from MiniBooNE appearance is subleading. The disappearance data which contribute to tension is from from CDHS \cite{Dydak:1983zq} and more recent experiments like IceCube \cite{Aartsen:2016xlq}, MINOS/MINOS+ \cite{Adamson:2011ku}, SK \cite{Abe:2014gda}, DeepCore \cite{Aartsen:2017bap}, MiniBooNE, NO$\nu$A \cite{Adamson:2017zcg}. 

There are several new experiments planned to test the sterile neutrino hypothesis\cite{Rott:2018rlw,Ko:2016owz,Serebrov:2018vdw,Haghighat:2018mve,Alekseev:2018efk,Tunnell:2012nu, Adey:2014rfv,Allemandou:2018vwb}. 
It was realized recently that beam-based long-baseline experiments 
can also probe the parameter space of the sterile neutrino models and several 
studies have been carried out in this direction considering the 
current as well as proposed experiments.   
Future experiments such as DUNE\cite{Acciarri:2015uup} or T2HK\cite{Abe:2016ero}  
are high statistics experiments and therefore the systematics are expected to 
play a crucial role, one of the major sources of systematic uncertainty are the 
neutrino-nucleus interaction cross-sections. 
Neutrinos from Stored Muons (nuSTORM)\cite{Tunnell:2012nu, Adey:2014rfv} is a facility proposed for the measurement of neutrino-nucleus cross-sections with percent-level precision. The high precision can be achieved because the stored-muon beam will allow the determination of neutrino flux with high accuracy. It has been shown that nuSTORM has excellent capability to 
search for the existence of light sterile neutrinos of the type postulated to explain the 
LSND and MiniBooNE results \cite{Athanassopoulos:1996jb, Aguilar:2001ty,Aguilar-Arevalo:2013pmq, Aguilar-Arevalo:2018gpe}. 

Beyond the Standard Model (BSM) physics descriptions have become essential in describing the non-zero neutrino mass after the discovery of neutrino oscillations. Non-zero neutrino masses can be generated by the ``see-saw" mechanism through an effective lepton number violating dimension-five operator of the form $LL\phi\phi$ which can be derived from physics beyond the Standard Model \cite{Weinberg:1979sa, Mohapatra:1979ia}. Such BSM physics can also lead to non-unitarity of the neutrino mixing matrix\cite{Goswami:2008mi,Rodejohann:2009cq,Malinsky:2009gw,Antusch:2006vwa,Antusch:2009pm,Malinsky:2009df,Blennow:2016jkn,Antusch:2014woa,Escrihuela:2015wra}. The unitarity of the PMNS matrix can be tested in accelerator-based neutrino oscillation experiments. Several studies have been performed to understand the implications of non-unitarity in present and future long baseline experiments \cite{Dutta:2016vcc,Miranda:2019ynh,deGouvea:2019ozk,Dutta:2019hmb}. In this context nuSTORM also holds promise for the study of the non-unitarity of the PMNS matrix and the constraint of the parameters which generate non-unitarity in the PMNS sector. 

The neutrino beam in nuSTORM originates from the muon decay process: 
$\mu^+ \rightarrow e^+\nu_e \overline{\nu}_\mu$ with 50\% $\nu_e$ and 50\% $\overline{\nu}_\mu$ which can give $e^-$ and 
$\mu^+$ at the detectors in absence of oscillation or any other new 
physics. If however there are flavour-changing processes then one can get 
wrong sign leptons which can constitute smoking-gun signals of new physics.  
A detector with charge identification capability is therefore ideal.

The sterile neutrino analysis performed in \cite{Behera:2016kwr} considered a magnetized iron-calorimeter 
detector with a superior  efficiency to identify the charge of the muons. 
This gives the detector the ability to record the 
$\mu^-$ events originating from $P_{\nu_e 
\nu_\mu}$ oscillations along with the 
 $\mu^+$ coming from the $P_{\overline{\nu}_\mu \overline{\nu}_\mu}$ 
channel.  In this analysis only the charged current events were considered. 
However, there are also a large number of neutral current (NC) events. In a three-flavour-mixing paradigm, given the flavour universality 
of the neutral current interactions and $P_{\mu e}+ P_{\mu \mu} + P_{\mu \tau} =1$, NC  events are not sensitive to the oscillation parameters. 
However, in the presence of new physics this may not be the case. 
For instance, for oscillations of muon neutrinos to a sterile neutrino, the rate of neutral-current events 
will be multiplied by $(1-P_{\mu s})$. 
The usefulness of NC events for sterile neutrino searches in the 
context of beam experiments has been studied in \cite{Abe:2019fyx,Adamson:2017zcg,Adamson:2011ku,Gandhi:2017vzo}. 

In this article we present the capabilities of nuSTORM in some sterile neutrino searches as well as the search for non-unitarity of the neutrino mixing matrix. In section:\ref{sec2} we discuss the nuSTORM proposal and the simulation of the facility. We discuss the results obtained in our study in section:\ref{sec3}, where subsection:\ref{sec3a} focuses on the study of sterile neutrinos at nuSTORM, while the consideration of the non-unitary of the neutrino mixing is presented in subsection:\ref{sec3b}. Conclusions are presented in section:\ref{conc}.

\section{Details of Simulation} \label{sec2}

We follow the configuration and detector simulations from \cite{Tunnell:2012nu,Adey:2014rfv,Evans:2013pka}. 
The unoscillated flux was taken from \cite{Tunnell:2012nu}. The simulation has been performed unsing the General Long Baseline Experiment Simulator (GLoBES) packege\cite{globes1, globes2}. 
The flux is based on the decay $\mu^+ \rightarrow e^+ + \nu_e + \bar{\nu}_{\mu} $.
The neutrino beam is generated with  
50 GeV protons with $2 \times 10^{21}$ protons on target over the duration of 10 years. Pions of 5 GeV are injected into the muon storage ring. Muons with energy of the order 3.8 GeV subsequently decay to give $\nu_e$ and $\bar{\nu}_{\mu} $.
 The $\nu_e$ flux peaks at 2.5 GeV whereas the $\bar{\nu}_{\mu}$ flux peaks at 3 GeV. nuSTORM is simulated as described in \cite{Tunnell:2012nu,Adey:2014rfv} . 
 
The primary aim for nuSTORM is the study of neutrino-nucleon scattering. At energies $E_\nu < 2~ \rm{GeV}$ quasi-elastic scattering and $1 \pi (\Delta)$ resonance are the dominant processes. But, at energies $E_\nu > 2~ \rm{GeV}$ the processes with multi-pion resonances along with shallow-and deep-inelastic scattering processes starts contributing significantly with deep-inelastic scattering process dominating at energies $E_\nu > 3~ \rm{GeV}$. These processes are not well understood yet. 
The nuSTORM facility \cite{nuSTORMatCERN} intends to study the interactions at these energies to understand these poorly-known processes. Therefore, to study the neutrino interactions at such wide range of energies the muon energy is expected to be between $1 < p_\mu < 6~ \rm{GeV/c}$. Additionally, the nuSTORM facility can also be optimized to study short-baseline oscillations with mass-squared difference $\Delta m^2_{\rm{LSND}} ~\sim~ 1~ \rm{eV^2}$ which requires an $L/E ~\sim~ 1~\rm{km}/\rm{GeV}$. This can be achieved by nuSTORM with the neutrino beam in the vicinity of $E_\nu ~\sim~2~\rm{GeV}$ and a baseline of 2 km.

\begin{figure}
\includegraphics[scale=0.5]{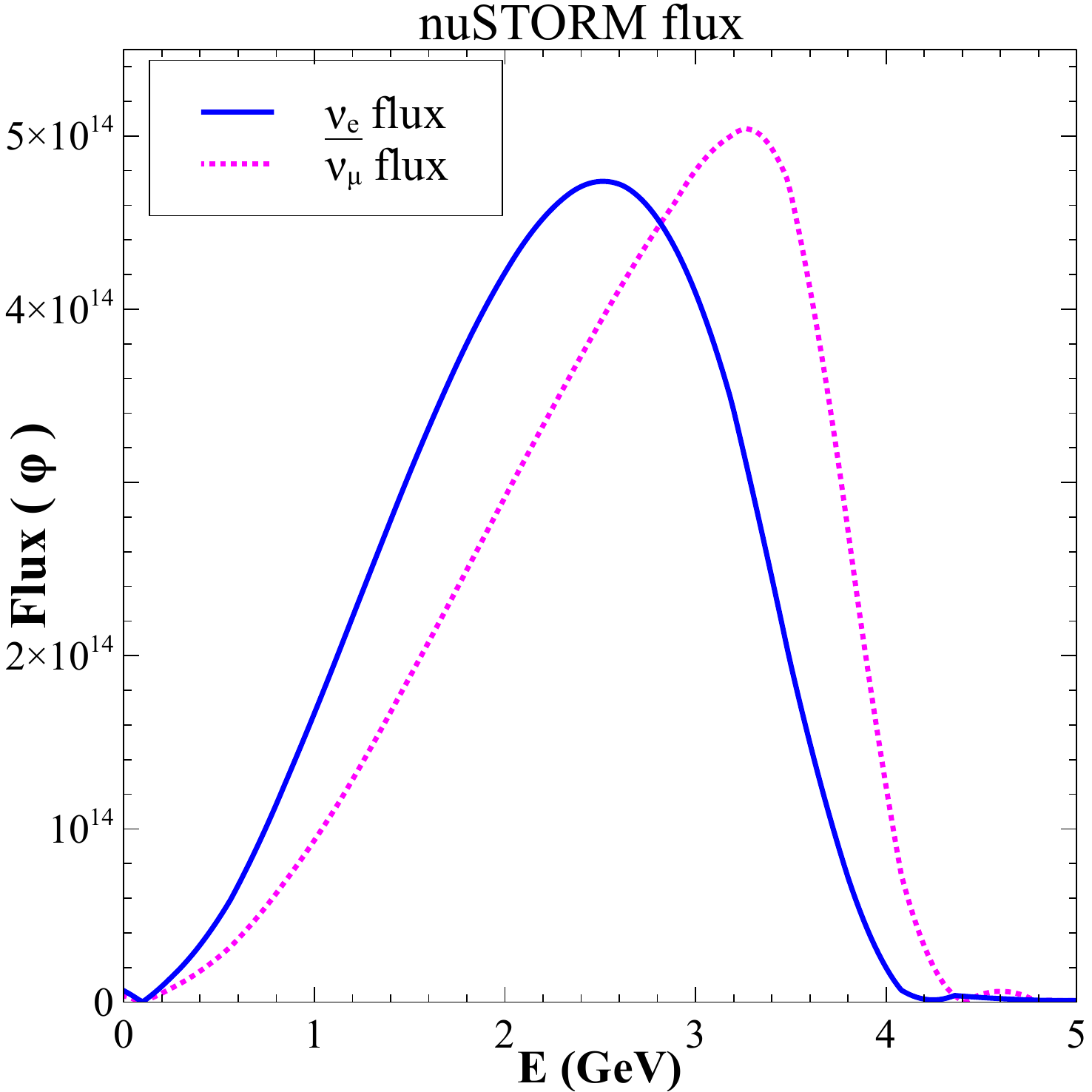}
\caption{The unoscillated $\nu_e$ and $\bar{\nu}_{\mu}$ flux extracted from the storage ring. The flux is evaluated for ($3.8 \pm 0.38$) GeV/c muon decay at a distance of 2 km\cite{Tunnell:2012nu}. }
\label{fig:flux}
\end{figure} 

%%%%%%%%%%%%%%%%%%%%%%%%%%%%%%%%%%%%%%%%%%%%%%%%%%%%%%%%%%%%%%%%%%%%%%%%%%%%%%%%%%%%%%%%%

In our simulation we consider a far detector at a distance of 2 km
from the source unless otherwise mentioned. 
The detector for the proposal has not yet been decided. In our case, following earlier studies \cite{Adey:2013pio} we chose a magnetized iron calorimeter detector because this detector can distinguish between $\nu_\mu$ and $\bar{\nu_\mu}$ so we can study $\nu_e \rightarrow \nu_\mu$ appearance channel as well as  $\bar{\nu_\mu} \rightarrow \bar{\nu_\mu}$ disappearance channel with the same beam. Alternatively, other detectors choices can also be explored in the future.
A 1.3 kt magnetized iron-scintillator calorimeter has
been selected as the detector for short-baseline oscillation
physics at nuSTORM as it has excellent charge selection
and detection characteristics for muons. The neutrino-nucleon scattering is the dominant interaction in the energy range of the nuSTORM flux. The important channels for this experiment are $\nu_e \rightarrow \nu_{\mu}$ appearance channel and $\bar{\nu}_{\mu} \rightarrow \bar{\nu}_{\mu}$ disappearance channel. 

The number of events in the $i^{th}$ energy bin
are calculated as 
\begin{eqnarray}
n^i_\alpha = \frac{N}{L^2} \int_{E_i - \frac{\Delta E_i}{2}}^{E_i + \frac{\Delta E_i}{2}} dE^\prime \int_{0}^{\infty} \varepsilon(E) \phi_\beta (E) P_{\alpha \beta} (E) \sigma_\alpha (E) R^c (E,E^\prime) \varepsilon^c (E^\prime) dE
\end{eqnarray}
where, $E$ denotes the true neutrino energy and $E^\prime$ denotes the measured neutrino energy.
$R^c (E,E^\prime)$ denotes the smearing matrix, which relates the true and 
the measured energy. This includes both kinematic smearing and the 
smearing due to energy reconstruction. 
This is often taken as a Gaussian. Migration matrices that give the probability for 
a neutrino generated in the $i^{th}$ energy bin to be reconstructed in the 
$j^{th}$ energy bin, if available from detector simulations, can also be used. 
$\varepsilon^c (E^\prime)$ denotes the post-smearing efficiency which 
contains, for instance, the information on energy cuts used. 
$\varepsilon(E)$ denotes the pre-smearing efficiency. 

In our analysis we have taken the energy resolution as a Gaussian . With
$R^{c}\left(E, E^{\prime}\right)=\frac{1}{\sigma(E) \sqrt{2 \pi}} e^{-\frac{\left(E-E^{\prime}\right)^{2}}{2 \sigma^{2}(E)}}$
and $\sigma(E)=0.15 E$.
The energy cuts are incorporated as ``post smearing efficiencies" 
as follows: 
\begin{eqnarray}
\epsilon^{c}\left(E^{\prime}\right) &=& 0~~~;~~~0~ {\rm GeV}~ <~E^{\prime}~<~1~ {\rm GeV}   \\
\epsilon^{c}\left(E^{\prime}\right) &=& 1~~~;~~~E^{\prime}~>~1~ {\rm GeV}
\end{eqnarray}

The event rates at the detector(multiplied by the efficiencies) 
and the corresponding pre-smearing efficiencies
are given in table \ref{tab:data1} and the unoscillated flux is given in fig.\ref{fig:flux}.

%%%%%%%%%%%%%%%%%%%%%%%%%%%%%%%%%%%%%%%%%%%%%%%%%%%%%%%%%%
\begin{table}[h!]
\begin{center}
\begin{tabular}{|c|c|c|}
\hline
Channel & $N_{events}$ & Efficiency at the detector \\
\hline \hline
$\nu_e \rightarrow \nu_{\mu}$ CC & 61 & 0.18 \\
\hline
$\nu_e \rightarrow \nu_e$ CC & 39865 & 0.18 \\
\hline
$\bar{\nu}_{\mu} \rightarrow \bar{\nu}_{\mu}$ NC & 8630 & 0.18 \\
\hline
$\bar{\nu}_{\mu} \rightarrow \bar{\nu}_{\mu}$ CC & 114983 & 0.94 \\
\hline
$\nu_e \rightarrow \nu_e$ NC & 13605 & 0.18 \\
\hline \hline
\end{tabular}
\caption{The events observed at the detector, this is equal to the expected number of events at the detector multiplied by their efficiencies according to \cite{Tunnell:2012nu,Adey:2014rfv}.}
\label{tab:data1}
\end{center}
\end{table}

The impact of the neutral current events is evaluated using a $\chi^2$ which is defined as
\begin{eqnarray}
\chi^2_{{\rm tot}} = \underset{\xi, \omega}{\mathrm{min}} \lbrace \sum_r ( \chi^2_{{\rm stat}}(\omega,\xi) + \chi^2_{{\rm pull}}(\xi) )_r  \rbrace.
%+ \chi^2_{prior}the
 \label{chi-tot}
\end{eqnarray}
r denotes the ``rules" and the statistical $\chi^2$ is $\chi^2_{{\rm stat}}$, systematic uncertainties are incorporated by $\chi^2_{pull}$ calculated
by the method of pulls with pull variables $\xi$.
The significance over each rule is calculated separately and the total $\chi^2$ is calculated by summation over all the various rules. Each ``rule" signifies a different channel .
The total $\chi^2$ is marginalized over the oscillation parameters.
The relevant oscillation parameters are represented by $\omega$.
%\{$\theta_{23},\theta_{12},\theta_{13}, \delta_{CP}, \Delta m^2_{21},\Delta m^2_{31},\theta_{14},\theta_{24},\theta_{34},\delta_{14}\delta_{24}$\}
%\begin{equation}\label{eq:total_chisq}
%\chi^{2} = \rm{Min} [ \chi^{2}_{stat}+\chi^{2}_{syst} ] .
%\end{equation}
The statistical $ \chi^{2}_{stat} $ is calculated assuming Poisson distribution,\begin{equation}\label{eq:stat_chisq}
 \chi^{2}_{stat} = \sum_i 2\left( N^{test}_i-N^{true}_i - N^{true}_i \log\dfrac{N^{test}_i}{N^{true}_i}\right).
\end{equation}
Here, `i' stands for the number of bins and $ N^{test}_i, N^{true}_i $ stands for
total number of test and true events respectively.
To include the effects of systematics in $ N^{test}_i$, the normalization and energy calibration errors are parametrized using the ``pull" and ``tilt" variables respectively. These are incorporated as follows:
%The method of pulls is taken into consideration to calculate the systematic uncertainties.
% The signal normalization uncertainties in case of DUNE(T2HK) are as follows:
%for $\nu_e$/$\bar{\nu_{e}}$ - 2\%(3.2\%) and $\nu_{\mu}$/$\bar{\nu_{\mu}}$ - 5\%(3.9\%). While the background uncertainties vary from 5\% to 20\% for DUNE and from 3.8\% to 5\% for T2HK.
\begin{eqnarray}
N^{{\rm (k)test}}_i(\omega, \xi) = \sum\limits_{k = s, b} N^{(k)}_{i}(\omega)[1 + c_{i}^{(k) norm} \xi^{(k) norm} + c_{i}^{(k) tilt} \xi^{(k) tilt} \frac{E_i - \bar{E}}{E_{max} - E_{min}}] \;,
\end{eqnarray}
where $k = s(b)$ represent the  signal (background) events. The effect of the pull variable $\xi^{norm}$(${ \xi}^{tilt}$) on the number of events is denoted by $c_i^{norm}$(${c_i}^{tilt}$). The bin-by-bin mean reconstructed energy is represented by $E_i$ where $i$ represents the bin. $E_{min}$, $E_{max}$ and $\bar{E} = ({E_{max} +E_{min}})/{2}$ are the minimum energy, maximum energy and the mean energy over this range.

\begin{center}
\begin{figure}[h!]
\hspace*{-2cm}
\includegraphics[height=6cm]{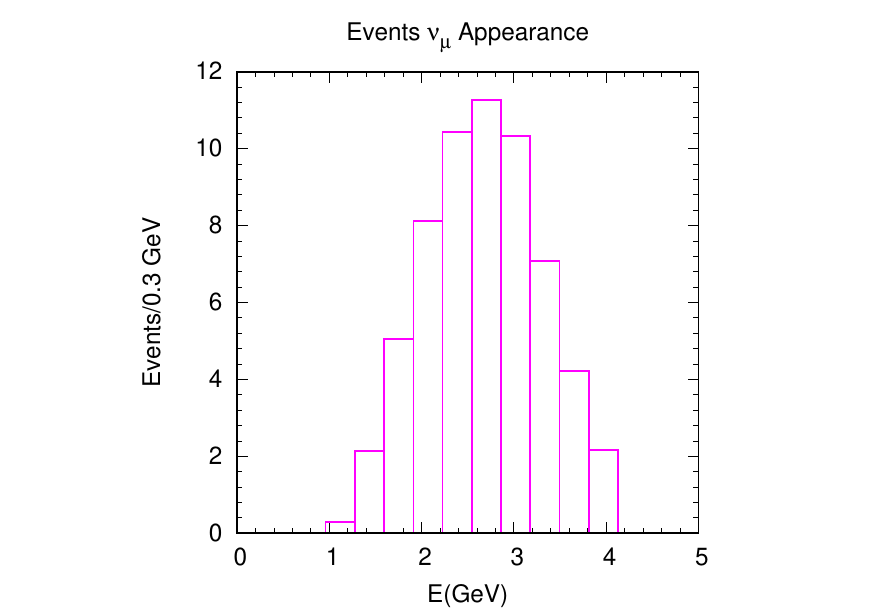}
\includegraphics[height=6cm]{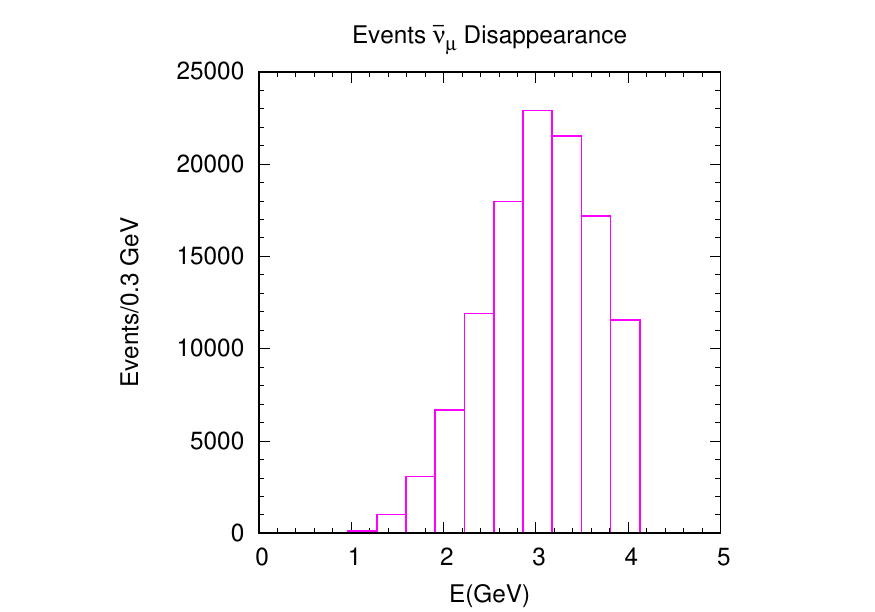}
\caption{The figure shows the distribution of events observed in absence of oscillation. A bin with bin width of 0.3 GeV. The left plot shows the appearance events while the right is for the disappearance events.}
\label{fig:events}
\end{figure} 
\end{center}

 The signal(background) normalization uncertainty for the appearance channel is 
taken as 1\%(10\%) \cite{Tunnell:2012nu,Adey:2014rfv} 
 while for $\nu_{\bar{\mu}}$ channel they are kept at 5\%(10\%). 
For NC the signal and background errors are taken to be 5\% and 10\% respectively.  
A background rejection factor of 
$ 10^{-3}$ is used for the disappearance channel while $ ~ 10^{-5}$ is used for appearance channel \cite{Tunnell:2012nu,Adey:2014rfv}. For NC events we use a background rejection factor of 
$ ~ 10^{-4}$. We have checked that the $\chi^2$ does not depend significantly 
on the background rejection factor for the NC analysis. The unoscillated events observed at the detector have been shown in the fig.\ref{fig:events}.

%%%%%%%%%%%%%%%%%%%%%%%%%%%%%%%%%%%%%%%%%%%%%%%%%%%%%%%%%%%%%%%%%%%%%%%%%%%%%%%%%%%%%%%%%

%%%%%%%%%%%%%%%%%%%%%%%%%%%%%%%%%%%%%%%%%%%%%%%%%%%%%%%%%%
\begin{table}[h!]
\begin{center}
\begin{tabular}{|c|c|c|}
\hline
Oscillation parameters & Value considered to simulate nuSTORM \\
\hline \hline
$\sin ^2 \theta_{13}$ & $0.022$ \\
\hline
$\sin ^2 \theta_{12}$ & $0.31$  \\
\hline
$\sin ^2 \theta_{23}$ & $0.558$ \\
\hline
$\Delta m^2_{21}$ (eV$^2$) & $7.39\times 10^{-5}$  \\
\hline
$|\Delta m^2_{31}|$ (eV$^2$) & $2.52\times 10^{-3}$   \\
\hline
$\delta$  & $0^\circ$  \\
\hline \hline
$\sin ^2 \theta_{14}$ & $0.025$ \\
\hline
$\sin ^2 \theta_{24}$ & $0.0.023$  \\
\hline
$\Delta m^2_{41}$ (eV$^2$) & $0.89$  \\
\hline \hline

\end{tabular}
\caption{The values of the 3 neutrino oscillation parameters \cite{nufit,Esteban:2018azc} and the representative values for 3+1 neutrino mixing \cite{Gariazzo:2017fdh} used in the present analysis. }
\label{tab:oscparam}  
\end{center}
\end{table}

Data is generated assuming the standard three-neutrino oscillations scenario as the null hypothesis and the new physics scenario under study is used as the alternative hypothesis.  
Schematically the number of events in the different channels can be written as 

\begin{eqnarray}
N^{CC}_\mu &=& \Phi(\nu_e) P_{e \mu} \sigma_{CC}  \\
N^{CC}_{\bar{\mu}} &=& \Phi(\bar{\nu_{\mu}}) P_{\bar{\mu} \bar{\mu}} \sigma_{CC} \\
%N^{NC}_{total} &=& \Phi(\bar{\nu_{\mu}}) P_{\bar{\mu} \bar{\mu}} \sigma_{NC} + \Phi(\nu_e) P_{e e} \sigma_{NC} \\
N^{NC}_{total} &=& \Phi(\bar{\nu_{\mu}}) (1 - P_{\mu s}) \sigma_{NC} +
 \Phi(\nu_e) (1-  P_{e s}) \sigma_{NC})
\end{eqnarray}

%\section{Probabilities} 

\section{Results and Discussions} \label{sec3}

\subsection{Sterile Neutrino} \label{sec3a}

Since we are considering a distance of $\sim$2 km and $E \sim$ 3 GeV there 
can be oscillations governed by a mass-squared difference of order $\rm{eV}^2$. Other terms do not contribute since the oscillation wavelengths are much larger. Thus we have the ``One Mass Scale Dominance" (OMSD) 
approximation in which the oscillation probabilities can be cast into 
an effective two flavor form.  
For the 3+1 picture, under the OMSD approximation, one has 

\be
P_{\alpha, \beta} =  4 |U_{\alpha 4}|^2 |U_{\beta 4}|^2 \sin^2 \left(  \frac{\Delta m_{41}^2 L}{4 E} \right)  
\ee
and, 
\be
P_{\alpha \alpha} = 1 - 4 |U_{\alpha 4}|^2 (1 - |U_{\alpha 4}|^2 ) \sin^2 \left(  \frac{\Delta m_{41}^2 L}{4 E} \right) 
\ee

\begin{center}
\begin{figure}[h!]
\hspace*{-2cm}
\includegraphics[height=6cm]{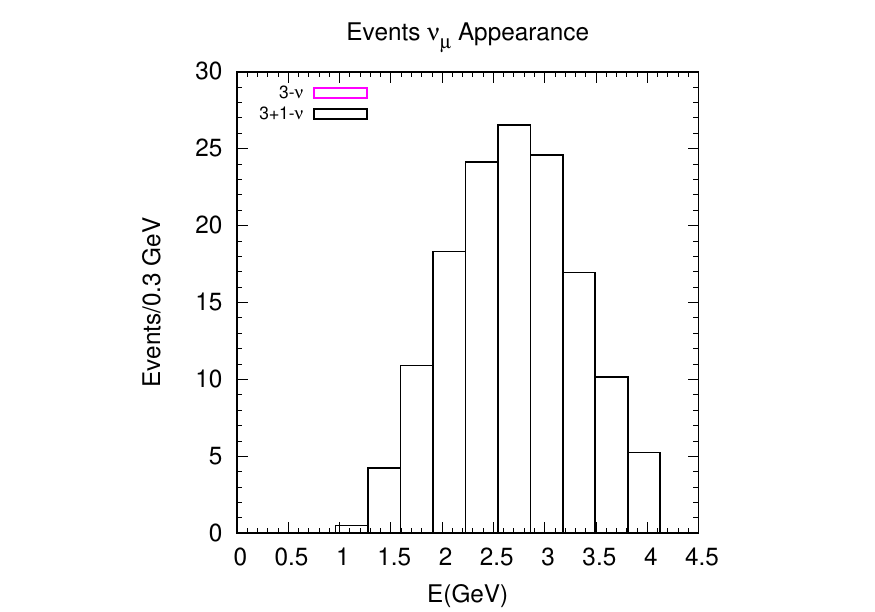}
\includegraphics[height=6cm]{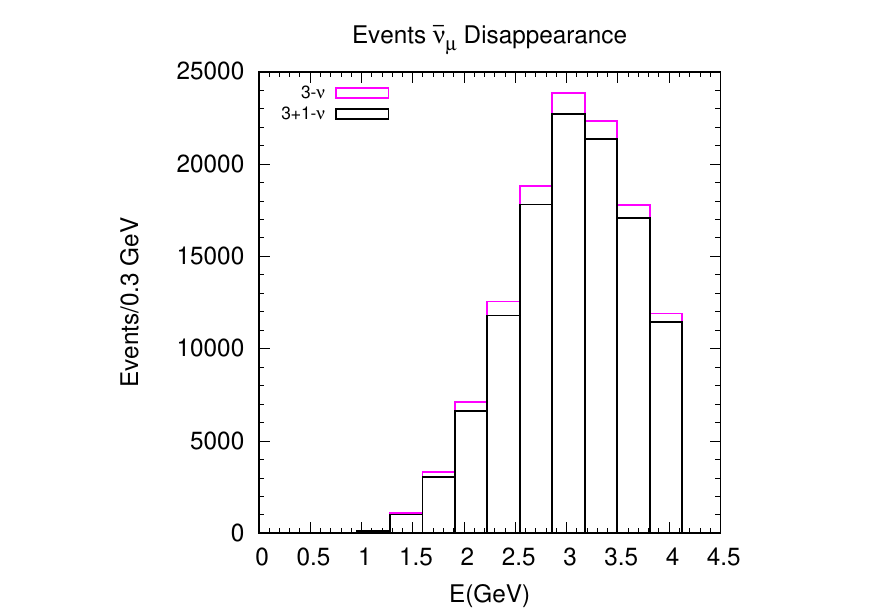}
\caption{The figure shows the distribution of events observed 3 neutrino and 3+1 neutrino scenarios. A bin with bin width of 0.3 GeV. The left plot shows the appearance events while the right is for the disappearance events. The magenta histograms are for events for standard three neutrino scenario while the black histograms are for 3+1 neutrino mixing.}
\label{fig:events-null-np}
\end{figure} 
\end{center} 

Bounds on individual mixing angles are derived using the parametrization 

\be \label{udef}
 U=R_{34}\tilde{R}_{24}\tilde{R}_{14}{R}_{23}\tilde{R}_{13}R_{12}.
\ee
Since we are in an effective two-generation  approximation, the phases do not appear in the oscillation probabilities and ignoring them one has, 

\begin{eqnarray}
U_{e4} & = & \sin\theta_{14} \nonumber 
\\
U_{\mu 4} & =  & \cos\theta_{14} \sin\theta_{24} \nonumber \\
U_{\tau 4} & = &  \cos\theta_{14} \sin\theta_{24} \nonumber \\
U_{s4} &  = &  \cos\theta_{14} \cos\theta_{24} \cos\theta_{34} 
\end{eqnarray} 

The relevant oscillation probabilities are given as 
\begin{eqnarray}
P_{e \mu} &=& 4 \cos^2 \theta_{14} \sin^2 \theta_{14} \sin^2 \theta_{24} \sin^2 \left(  \frac{\Delta m_{41}^2 L}{4 E} \right) \label{eq:p_emu} \\
P_{\mu \mu} &=& 1 - 4 \sin^2 \theta_{24} \cos^2 \theta_{14}(1 - \sin^2 \theta_{24} \cos^2 \theta_{14}) \sin^2 \left(  \frac{\Delta m_{41}^2 L}{4 E} \right)\label{eq:p_mumu} \\
P_{\mu s} &=& 4 \cos^4 \theta_{14} \cos^2 \theta_{24} \cos^2 \theta_{34} \sin^2 \theta_{24} \sin^2 \left(  \frac{\Delta m_{41}^2 L}{4 E} \right)\label{eq:p_mus} \\
%P_{e e} &=& 1 - 4 \sin^2 \theta_{14} \cos^2 \theta_{14} \sin^2 \left(  \frac{\Delta m_{41}^2 L}{4 E} \right) \\
P_{e s} &=& 4 \cos^2 \theta_{14} \sin^2 \theta_{14} \cos^2 \theta_{24} \cos^2 \theta_{34} \sin^2 \left(  \frac{\Delta m_{41}^2 L}{4 E} \right)\label{eq:p_es}
\end{eqnarray}

The comparison between the event spectrum three neutrino scenarios and 3+1 neutrino mixing is shown in fig.\ref{fig:events-null-np} for appearance channel and disappearance channel. In case of the appearance channel the oscillations due to $\Delta m^2_{31}$ are yet to develop so there are no events for three neutrino mixing. But, such short baselines are enough to develop oscillations due to $\Delta m^2_{41} \sim 1~ \rm{eV^2}$. Therefore, the appearance flux is non-zero in case of 3+1 neutrino mixing, which makes such experiments ideal for new physics searches. The same reason is also valid for the disappearance channel where the flux in case of 3+1 neutrino mixing is less compared to that of three neutrino mixing.

\begin{equation}
\sin^2 2\theta_{\mu e} = 4 |U_{e4}|^2 |U_{\mu 4}|^2 = 4 s_{14}^2 c_{14}^2 s_{24}^2\label{eq:th_mue}
\end{equation}
\begin{equation}
\sin^2 2\theta_{\mu \mu} = 4 |U_{\mu 4}|^2 (1 - |U_{\mu 4}|^2) = 4 c_{14}^2 
s_{24}^2 (1 - c_{14}^2 s_{24}^2)\label{eq:th_mumu}
\end{equation}  
 
%The effect of inclusion of NC events on the appearance and disappearance channels are shown using the exclusion plots. 

\begin{center}
\begin{figure}[h!]
\hspace*{-1.0cm}
\includegraphics[height=6cm]{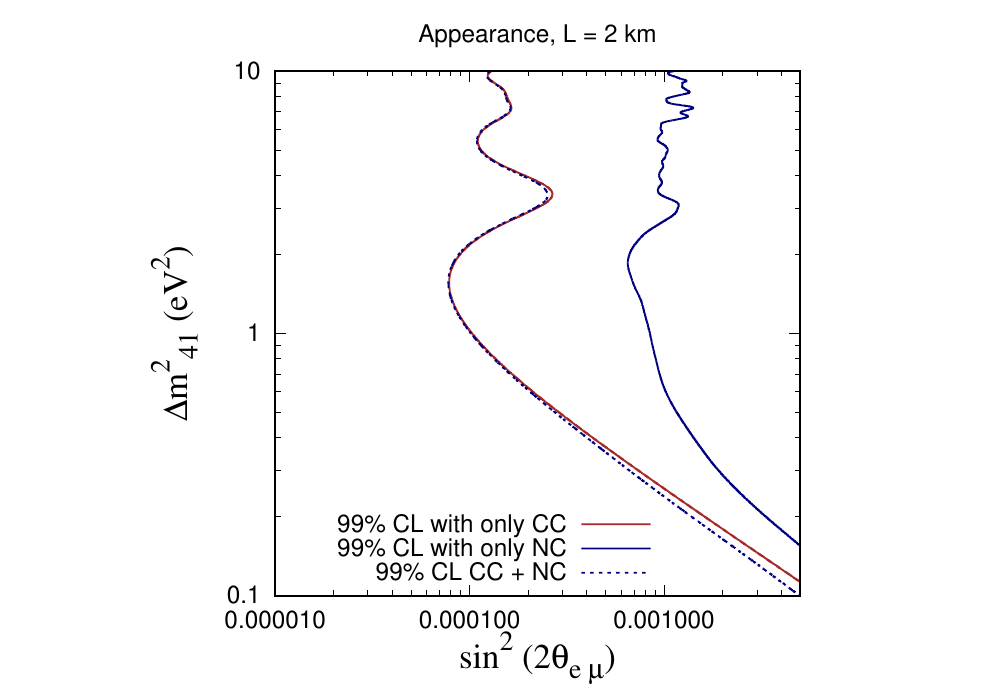}
\hspace*{-1.0cm}
\includegraphics[height=6cm]{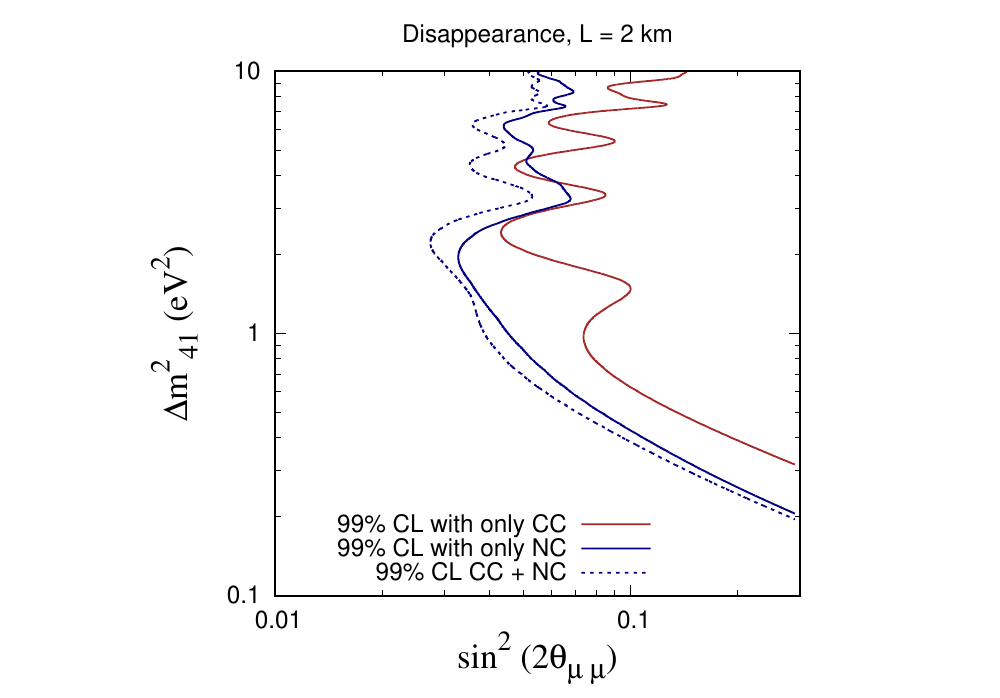} \\
\hspace*{-1.0cm}
\includegraphics[height=6cm]{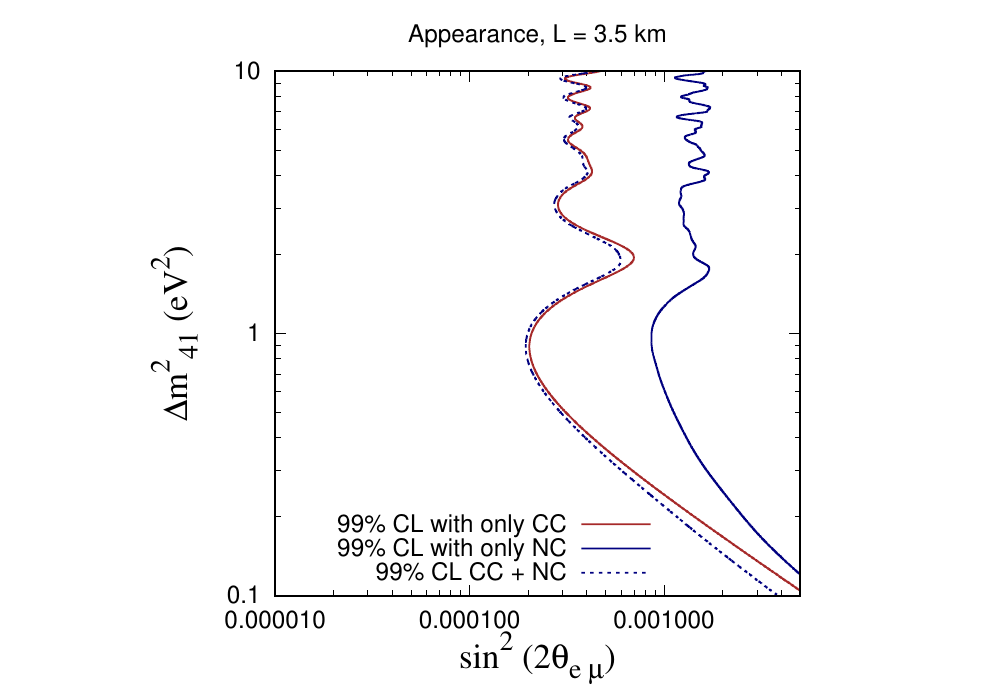}
\hspace*{-1.0cm}
\includegraphics[height=6cm]{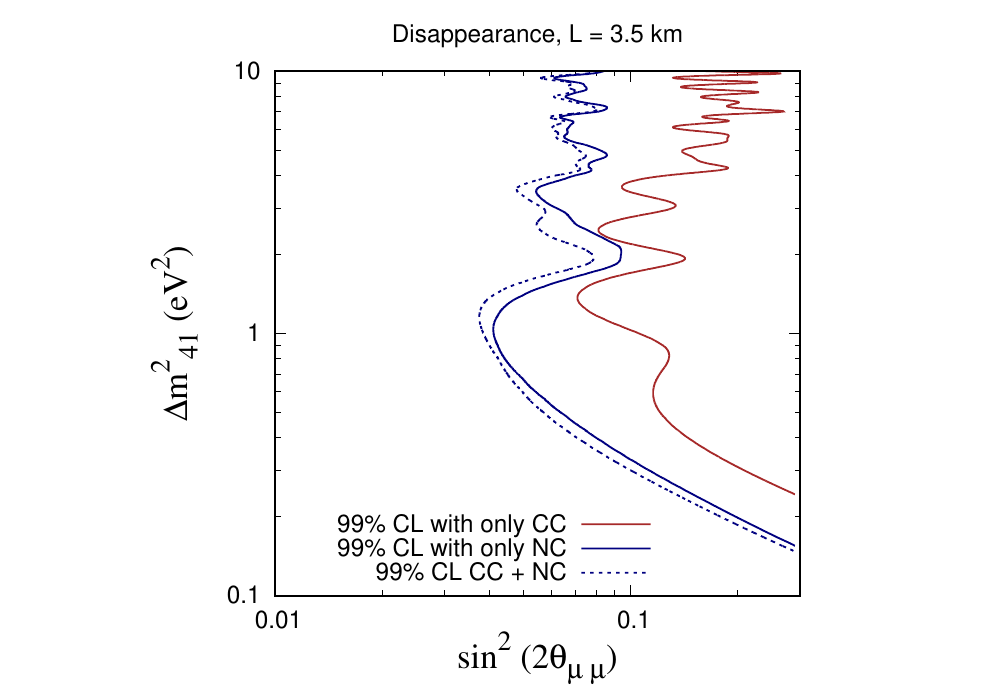}
\caption{The testable regions for sterile neutrinos as predicted by nuSTORM in terms of $\Delta m^2_{41}$ vs $\sin^2 \theta_{\mu e}$ for the left plots and $\Delta m^2_{41}$ vs $\sin^2 \theta_{\mu \mu}$ for the right. The first row indicates the sensitivities or baseline of 2 km while the second row for 3.5 km.
Each plot consists of 3 contours of 99\% confidence level significance exclusion regions for various channels as labeled in the plots.}
\label{fig:dm41-th-bounds}
\end{figure} 
\end{center}

Figure \ref{fig:dm41-th-bounds} shows the bounds on $\Delta m^2_{41}$ with respect to the effective mixing angles $\theta_{\mu e}$ and $\theta_{\mu \mu}$ for baselines of 2 km and 3.5 km. The oscillation amplitudes satisfy: $P_{e \mu} \propto s_{14}^2 s_{24}^2$; $P_{\mu \mu} \propto 1 - s_{24}^2$; and $P_{e s} + P_{\mu s} \propto s_{14}^2 + s_{24}^2$ (see eq.\ref{eq:p_emu} - \ref{eq:p_es}). Therefore, in the case of the appearance channel, $P_{e \mu}$ can constrain the effective mixing angle $\sin^2 2\theta_{\mu e}$ which is a product of $s_{14}^2 s_{24}^2$ while the neutral-current channel cannot constrain the product of $s_{14}^2 s_{24}^2$ and hence cannot efficiently constrain $\sin^2 2\theta_{\mu e}$. The disappearance channel effectively probes the parameter $\theta_{\mu \mu}$ in terms of the parameter $s_{24}^2$, also the neutral current channel probes $s_{14}^2 + s_{24}^2$, hence, the neutral current channel can significantly constrain the parameter $\theta_{\mu \mu}$. From \cite{Diaz:2019fwt} we can see that the current global bes-fit in  $\Delta m^2_{41} - \sin^2 2\theta_{\mu e}$ plane lies around $\Delta m^2_{41} \sim 1~\rm{eV^2}$ and $\sin^2 2\theta_{\mu e} \sim 10^{-3}$ so, nuSTORM has the capability to test the current best-fit and can constrain the parameter space further up to $\sin^2 2\theta_{\mu e} \sim 10^{-4}$. The current best-fit for $\Delta m^2_{41} - \sin^2 2\theta_{\mu \mu}$ plane which is $\Delta m^2_{41} \sim 1~\rm{eV^2}$ and $\sin^2 2\theta_{\mu \mu} \sim 10^{-1}$ also lies within the testable region of the nuSTORM experiment. Therefore, nuSTORM can not only test the current allowed parameter space but can also put further constrains on the currently allowed parameter space. 
%This study was performed for two baselines of 2 km and 3.5 km. 
The 2 km baseline was chosen for comparison to the results presented in \cite{LAGRANGE20161771}. The choice of the 3.5 km baseline was motivated by the fact that this places the detector at oscillation maxima for $\Delta m^2_{41} ~\sim~ 1 \rm{eV^2}$. If we study the bottom panel of the fig.\ref{fig:dm41-th-bounds} we observe that the best sensitivities for both $\theta_{\mu e}$ and $\theta_{\mu \mu}$ are observed around $\Delta m^2_{41} ~\sim~ 1 \rm{eV^2}$, which is expected. Proceeding to the top panel of the fig.\ref{fig:dm41-th-bounds} we find that the most sensitive region has shifted to $\Delta m^2_{41} ~\sim~ 1.2 \rm{eV^2}$, this is expected because $\Delta m^2_{41} L \approx 3.7 \rm{eV}^2 \rm{km}$. However, the overall sensitivity is better for the lower baseline of 2 km as the lower baseline has a lower statistical uncertainty because of a higher flux at the detector.

%The inclusion of NC events do not affect the appearance channel 
%so much since the appearance channel has a higher sensitivity than the 
%NC. But since disappearance channel and NC have the same order of sensitivities the 
%inclusion of NC has more effect on the disappearance channel.  

%\begin{tabular}{cl}
%\begin{tabular}{c}
%\parbox{0.4\linewidth}{
%\includegraphics[height=6.5cm,width=6cm]{delm41-bound-nustorm-sterile-wnc-th14-th24-1.pdf}
%}
%\end{tabular}
%& \begin{tabular}{l}
%\parbox{0.6\linewidth}{
%\includegraphics[height=3cm,width=6cm]{delm41-bound-nustorm-sterile-wnc-th14-th34-1.pdf} \\
%\includegraphics[height=3cm,width=6cm]{delm41-bound-nustorm-sterile-wnc-th24-th34-1.pdf}
%%\caption{The testable regions for sterile neutrinos as predicted by nuSTORM in terms of $\theta_{14}$, $\theta_{24}$ and $\theta_{34}$ bounds. The first, second and third plots present the $\theta_{14}$(test) vs $\theta_{24}$(test), $\theta_{14}$(test) vs $\theta_{34}$(test) and $\theta_{24}$(test) vs $\theta_{34}$(test) contours respectively. Each plot consists of 5 contours of 99\% confidence level significance exclusion regions for various channels as labeled in the plots.  }
%}
%\end{tabular}
%\end{tabular} 

\begin{center}
\begin{figure}[h!]
\hspace*{-2cm}
\includegraphics[height=10cm,width=1.3\textwidth]{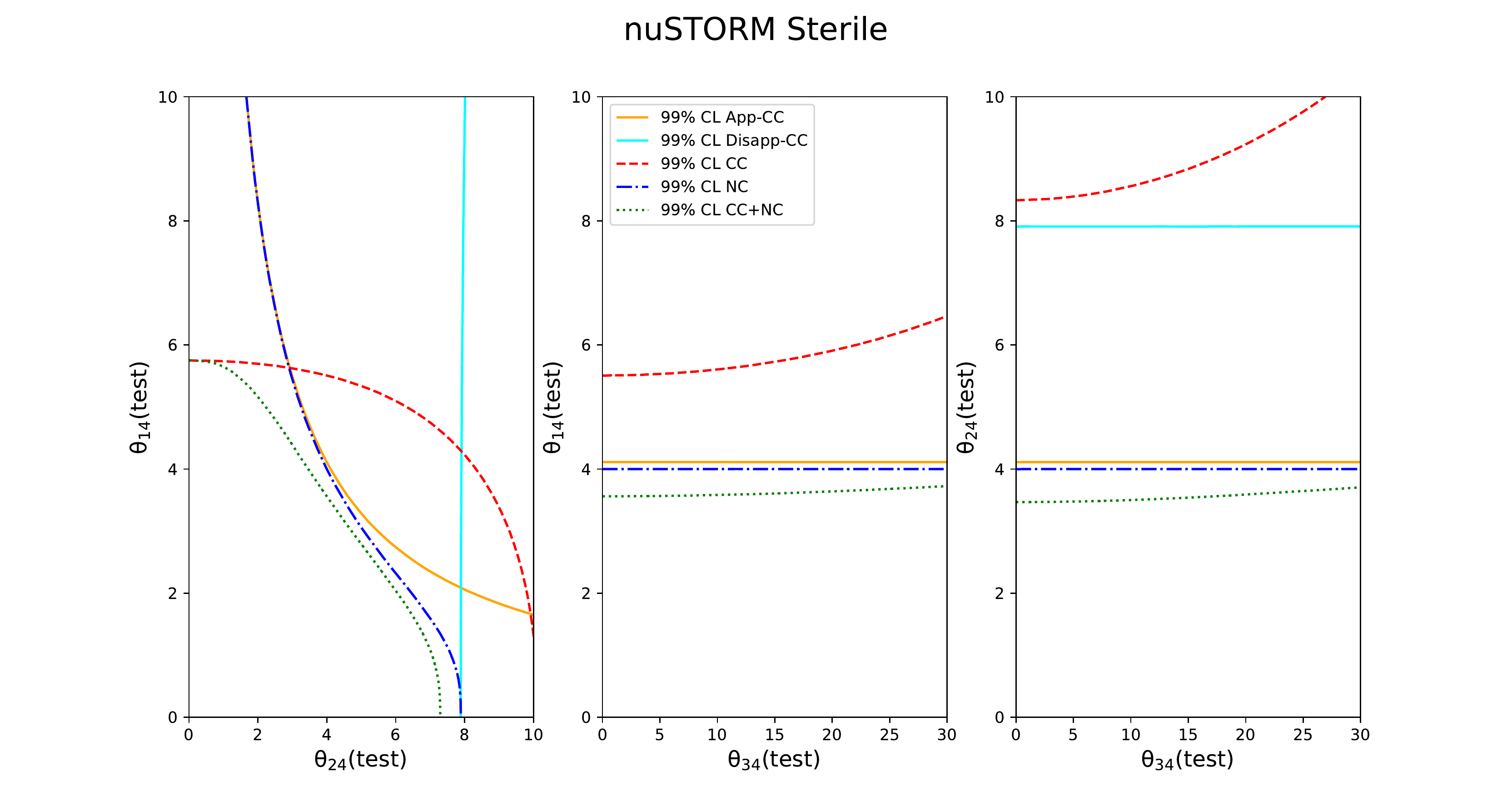}
%\includegraphics[height=6cm]{event-sterile-nc-th14.pdf} \\
%\end{figure}
%\end{center}
%%
%%\begin{center}
%%\begin{figure}[h!]
%%\hspace*{-0.5cm}
%\includegraphics[height=4cm]{delm41-bound-nustorm-sterile-wnc-th14-th34-1.pdf}
%%\includegraphics[height=6cm]{event-sterile-nc-th24.pdf}
%%\hspace*{-0.5cm}
%\includegraphics[height=4cm]{delm41-bound-nustorm-sterile-wnc-th24-th34-1.pdf}
\caption{The testable regions for sterile neutrinos at nuSTORM for $\Delta m^2_{41} = 1\rm{ev^2}$ and baseline of 2 km in terms of $\theta_{14}$, $\theta_{24}$ and $\theta_{34}$ bounds. Here,  $\theta_{14}$, $\theta_{24}$ and $\theta_{34}$ are in degrees. The first, second and third plots present the $\theta_{14}$(test) vs $\theta_{24}$(test), $\theta_{14}$(test) vs $\theta_{34}$(test) and $\theta_{24}$(test) vs $\theta_{34}$(test) contours respectively. Each plot consists of 5 contours of 99\% confidence level significance exclusion regions for various channels as labeled in the plots.  }
\label{fig:theta-bounds}
\end{figure}
\end{center}

Figure \ref{fig:theta-bounds} presents the predicted $\theta_{14}$, $\theta_{24}$ and $\theta_{34}$ bounds expected from nuSTORM. The first plot from fig.\ref{fig:theta-bounds} shows the $\theta_{14}$ versus $\theta_{24}$ exclusion region considering the data generated from 3 flavour oscillation with parameters as given in tab:\ref{tab:data1}, but setting the fourth generation parameters to zero. The solid orange line shows the $\theta_{14}$ versus $\theta_{24}$ exclusion region predicted from the appearance channel, the relevant probability for this channel is $P_{e \mu}$ given by the expression in eq:\ref{eq:p_emu}. As, the allowed regions for $\theta_{14}$, $\theta_{24}$ are small hence the expression for $P_{e \mu}$ at constant energy and baseline is roughly  $\propto \theta^2_{14} \theta^2_{24}$ which explains the hyperbolic nature of the charged current appearance plot. The disappearance probability $P_{\bar{\mu} \bar{\mu}}$ approximately reduces to $1-4\theta^2_{24}$, which is independent of $\theta_{14}$, so $\theta_{14}$ remains unaffected by the disappearance channel. Another important channel which can be probed is the neutral-current channel. The total contribution to the neutral-current channel comes from $P_{\mu s} + P_{e s}$ because neutral-current events from neutrino and antineutrino cannot be differentiated by the detector. The total neutral-current probability approximately reduces to $P_{\mu s} + P_{e s} \propto \theta^2_{14} + \theta^2_{24}$, which describes the approximate elliptical nature of the neutral current channel given by red dashed lines in the fig:\ref{fig:theta-bounds}. The total CC event curve(blue dashed curve) is the total contribution of appearance CC and disappearance CC. While the green dotted curve presents the contribution of all the above channels i.e. the total CC and NC event samples. It is clear from the figure that the inclusion of NC events can put stringent bounds on both $\theta_{14}$ and $\theta_{24}$. 
%{\bf Moreover, without NC $\theta_{24}$ was unbounded therefore the interesting result is inclusion of NC events can give an upper bound on $\theta_{14}$. 
We can conclude from this study that nuSTORM will be able to test $\theta_{14}$, $\theta_{24}$ up to $6^\circ$ and $7.5^\circ$ respectively. Comparing the results obtained with the expected sensitivity of DUNE \cite{Coloma:2017ptb} it was found that neutral current events from DUNE can resolve $\theta_{14}$ up to $10^\circ$ and $\theta_{24}$ upto $15^\circ$ with $5\%$ systematics for $\Delta m^2_{41} = 0.5 \rm{eV^2}$.

The second and third plots in the figure show the ability of nuSTORM to constrain $\theta_{14}$ and $\theta_{24}$ with respect to $\theta_{34}$. Taking all the channels into account both $\theta_{14}$ and $\theta_{24}$ can be approximately constrained up to $4^\circ$ at nuSTORM. In both the plots it is clear that the charged current interactions are independent of $\theta_{34}$ which is also understood from the expressions for $P_{e \mu}$ and $P_{\bar{\mu} \bar{\mu}}$. The only dependence on $\theta_{34}$ can come from the neutral current channel. However, $P_{e \mu} + P_{\bar{\mu} \bar{\mu}} \propto \cos^2 \theta_{34}$, as a result of which there is weak dependence of $\theta_{34}$ on the neutral current events hence $\theta_{34}$ cannot be constrained by neutral current events in nuSTORM. 

\begin{center}
\begin{figure}[h!]
\hspace*{-2cm}
\includegraphics[height=10cm,width=1.3\textwidth]{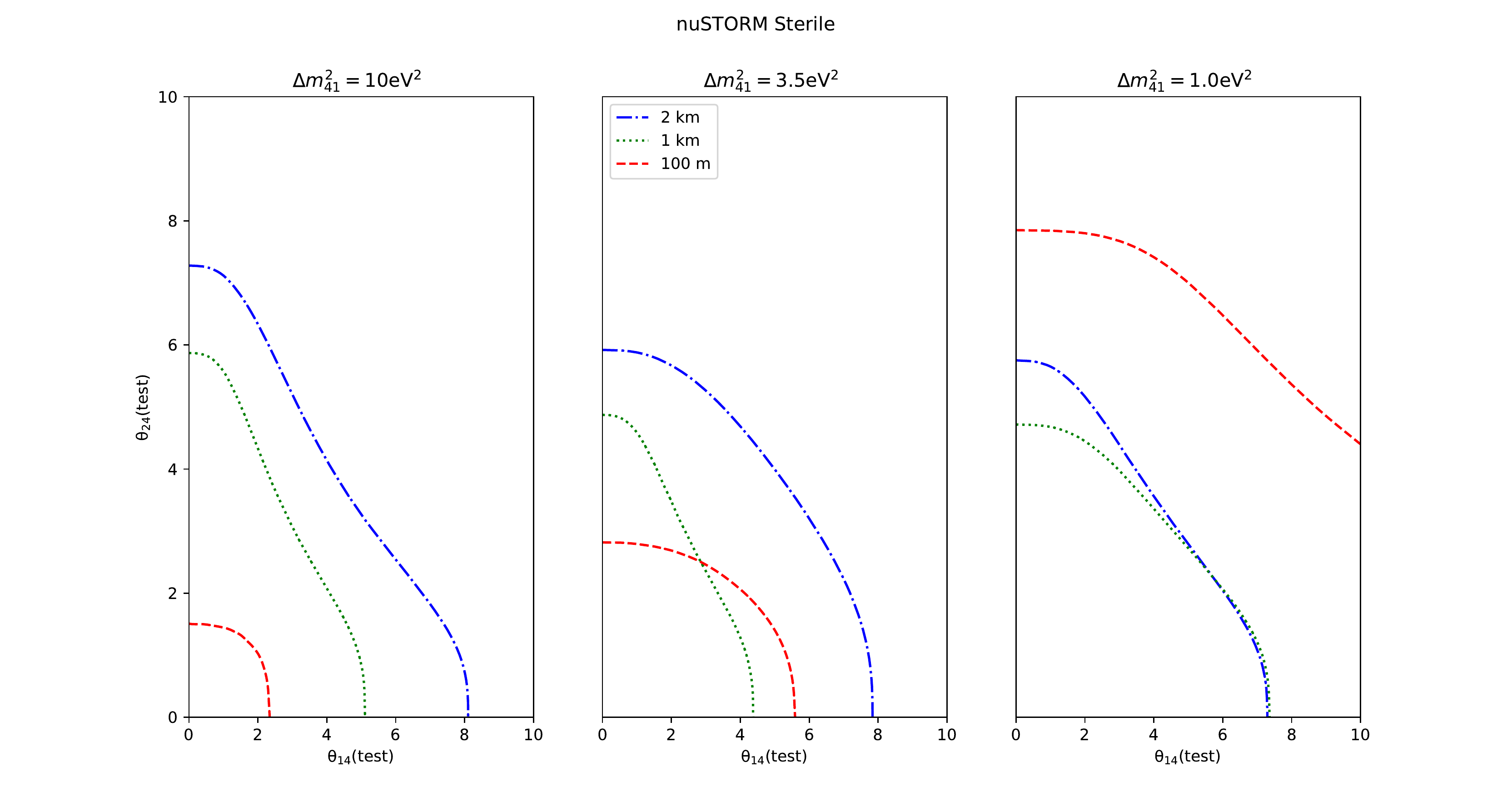}
\caption{The testable regions for sterile neutrinos at nuSTORM in terms of $\theta_{14}$ vs $\theta_{24}$ bounds. Here, $\theta_{14}$, $\theta_{24}$ are in degrees. The first plot presents the $\theta_{14}$(test) vs $\theta_{24}$(test) for $\Delta m^2_{41} = 1 \rm{eV}^2$,the second plot for $\Delta m^2_{41} = 3.5 \rm{eV}^2$ and the third plot for $\Delta m^2_{41} = 10 \rm{eV}^2$. Each plot consists of 3 contours of 99\% confidence level significance exclusion regions for various baselines as labeled in the plots.}
\label{fig:theta-baseline}
\end{figure}
\end{center}
The left plot in fig:\ref{fig:theta-baseline} shows the effect of varying the baseline of nuSTORM on the bounds in the $\theta_{14}$, $\theta_{24}$ plane for $\Delta m^2_{41} = 1 \rm{eV}^2$. The best sensitivity of an experiment is observed at the oscillation maxima. The first oscillation maximum is given by $1.27 \Delta m^2_{41} L / E = \pi / 2$. As the mean energy of the experiment is $\sim 3~ \rm{GeV}$, $\Delta m^2_{41} L \approx 3.7 \rm{eV}^2~ \rm{km}$. It is evident from the relation that probing a larger $\Delta m^2_{41}$ requires a smaller baseline($L$) and vice versa. Analyzing the red curves in the fig:\ref{fig:theta-baseline}, which show result for a the baseline of 100 m, we observe that as $\Delta m^2_{41}$ is increased the sensitivity also increases. Similarly, if we observe the  green curves representing a 1 km baseline, we observe that the best sensitivity is obtained for the case $\Delta m^2_{41} = 3.5~ \rm{eV}^2$ which is expected from the above relation. Deviation from $\Delta m^2_{41} = 3.5 \rm{eV}^2$, on either side compromises the sensitivity. The blue curves demonstrate the sensitivities of the 2 km baseline for nuSTORM. The $\Delta m^2_{41} \approx 1.8 \rm{eV}^2~ \rm{km}$ is expected to have the best sensitivity for the 2 km baseline. As we increase the $\Delta m^2_{41}$ gradually the sensitivity decreases with increasing $\Delta m^2_{41}$. We observe that the 1 km baseline has good sensitivity for both $\theta_{14}$ and $\theta_{24}$ consistently over the range of $\Delta m^2_{41}$.

%
%\begin{center}
%\begin{figure}[h!]
%\hspace*{-2cm}
%\includegraphics[height=6cm]{delm41-bound-nustorm-sterile-wnc-th24-th34-1.pdf}
%\includegraphics[height=6cm]{event-sterile-nc-th34.pdf}
%%\hspace*{10cm}
%\end{figure}
%\end{center}

%\begin{figure}[h!]
%\hspace*{-2cm}
%\includegraphics[height=7cm]{delm41-bound-nustorm-sterile-appearance-wnc-glb1-min-corrected2-min.eps}
%\includegraphics[height=7cm]{delm41-bound-nustorm-sterile-disappearance-wnc-glb1-min-corrected2-min.eps}
%\end{figure}

\subsection{Non-Unitarity} \label{sec3b}
In presence of non unitarity, the time evolution of the mass eigenstate in vacuum is:
\begin{eqnarray}
i\dfrac{d \mid \nu_i \rangle}{dt} =  H \mid \nu_i \rangle,
\end{eqnarray}
where $H$ is the Hamiltonian in the mass basis.
After time t($\equiv$L), the flavour state can be written as 
\begin{eqnarray}
|\nu_{\alpha}(t)\rangle =  N^{*}_{\alpha i} |\nu_i (t)\rangle =  N^{*}_{\alpha i}(e^{-iHt})_{ij}|\nu_j (t=0)\rangle.
\end{eqnarray}
%Therefore, the Hamiltonian can be expressed as:
%\begin{equation}
%\label{eq:H-nu}
%H = \frac{1}{2E}\begin{pmatrix}
%0 & 0 & 0 \\
%0 & \Delta m_{21}^2 & 0 \\
%0 & 0 & \Delta m_{31}^2
%\end{pmatrix} + N^\dagger  \begin{pmatrix}
%V_{\rm CC}+V_{\rm NC} & 0 & 0 \\
%0 & V_{\rm NC} & 0 \\
%0 & 0 & V_{\rm NC}
%\end{pmatrix}  N,
%\end{equation}
In this framework the mixing matrix $N$ can be parametrized as:
\begin{eqnarray}
 N = N^{ NP}  U = \begin{bmatrix}
    \alpha_{11} & 0 &0 \\
    \alpha_{21} & \alpha_{22} & 0\\
    \alpha_{31} & \alpha_{32} & \alpha_{33}
  \end{bmatrix} U;
\end{eqnarray}
where $U$ is the PMNS matrix, $N^{ NP}$ is the left triangle matrix which parametrizes the non unitarity. In the matrix $N^{ NP}$ the diagonal elements are real and the off diagonal elements can be complex.  

The above discussions leads us to the transition probability:
\begin{eqnarray}
\label{eq:nu_tranprob}
P(\nu_{\alpha} \rightarrow \nu_{\beta}) = |\langle\nu_{\beta}|\nu_{\alpha}(t)\rangle|^2=|N^{\ast}_{\alpha i}diag(e^{-i\Delta m^{2}_{i1}t/2E})_{ij} N_{\beta j}|^2
\end{eqnarray}

Using the above parametrization the transition probabilities $P_{\mu e}$ and $P_{\mu \mu}$ can be written:
\begin{eqnarray}
P_{e \mu} = \alpha_{11}^2 |\alpha_{21}|^2  &-& 4  
        \sum^3_{j>i} Re\left[ 
        N^*_{\mu j}N_{ej}N_{\mu i}N^*_{ei} \right]
        \sin^2\left(\frac{\Delta m^2_{ji}L}{4E}\right)  \nonumber \\
&+& 
        2 \sum^3_{j>i} Im\left[
        N^*_{\mu j}N_{ej}N_{\mu i}N^*_{ei}\right] 
        \sin\left(\frac{\Delta m^2_{ji}L}{2E}\right)  
        . 
\end{eqnarray}

\begin{equation}
P_{\mu\mu} =  (|\alpha_{21} |^2 + \alpha_{22}^2)^2 - 
4\sum^3_{j>i} |N_{\mu j}|^2|N_{\mu i}|^2 \sin^2\left(\frac{\Delta m^2_{ji}}{4E}L\right).
\end{equation}
For nuSTORM, with a baseline of 2 km, the transition probabilities become independent of the baseline length because $\frac{\Delta m^2 L}{E} \ll 1$. Therefore, the relevant transition probabilities are:

\begin{eqnarray}
P_{e \mu} &=& \alpha_{11}^2 |\alpha_{21}|^2, \rm{and} \\
P_{\mu \mu} &=&  (|\alpha_{21} |^2 + \alpha_{22}^2)^2
\end{eqnarray} 

Along with the charged current events, neutral-current events can also be helpful in studying the non-unitarity of the mixing matrix. 
The important probabilities for the inclusion of the neutral current events are
\begin{eqnarray}
P_{e s} &=& 1-(\alpha_{11}^2 (\alpha_{11}^2+|\alpha_{21}|^2+|\alpha_{31}|^2)); \rm{and} \\
P_{\mu s} &=&  1-(\alpha_{11}^2 |\alpha_{21}|^2+\alpha_{22}^4 + 2\alpha_{22}^2 |\alpha_{21}|^2 +\alpha_{22}^2 |\alpha_{32}|^2) 
\end{eqnarray} 
The detector cannot distinguish the various kinds of neutral current events, so we can probe the total neutral current probability: 
\begin{eqnarray}
P_{e s}+P_{\mu s} &=& 2- (\alpha_{11}^2 (\alpha_{11}^2+2|\alpha_{21}|^2+|\alpha_{31}|^2)+\alpha_{22}^2 (\alpha_{22}^2+2|\alpha_{21}|^2+|\alpha_{32}|^2)).
%P_{e s}+P_{\mu s} &\approx & 2-(\alpha_{11}^2+\alpha_{22}^2)
\end{eqnarray}

\begin{center}
\begin{figure}[h!]
\hspace*{-2cm}
\includegraphics[height=10cm,width=1.3\textwidth]{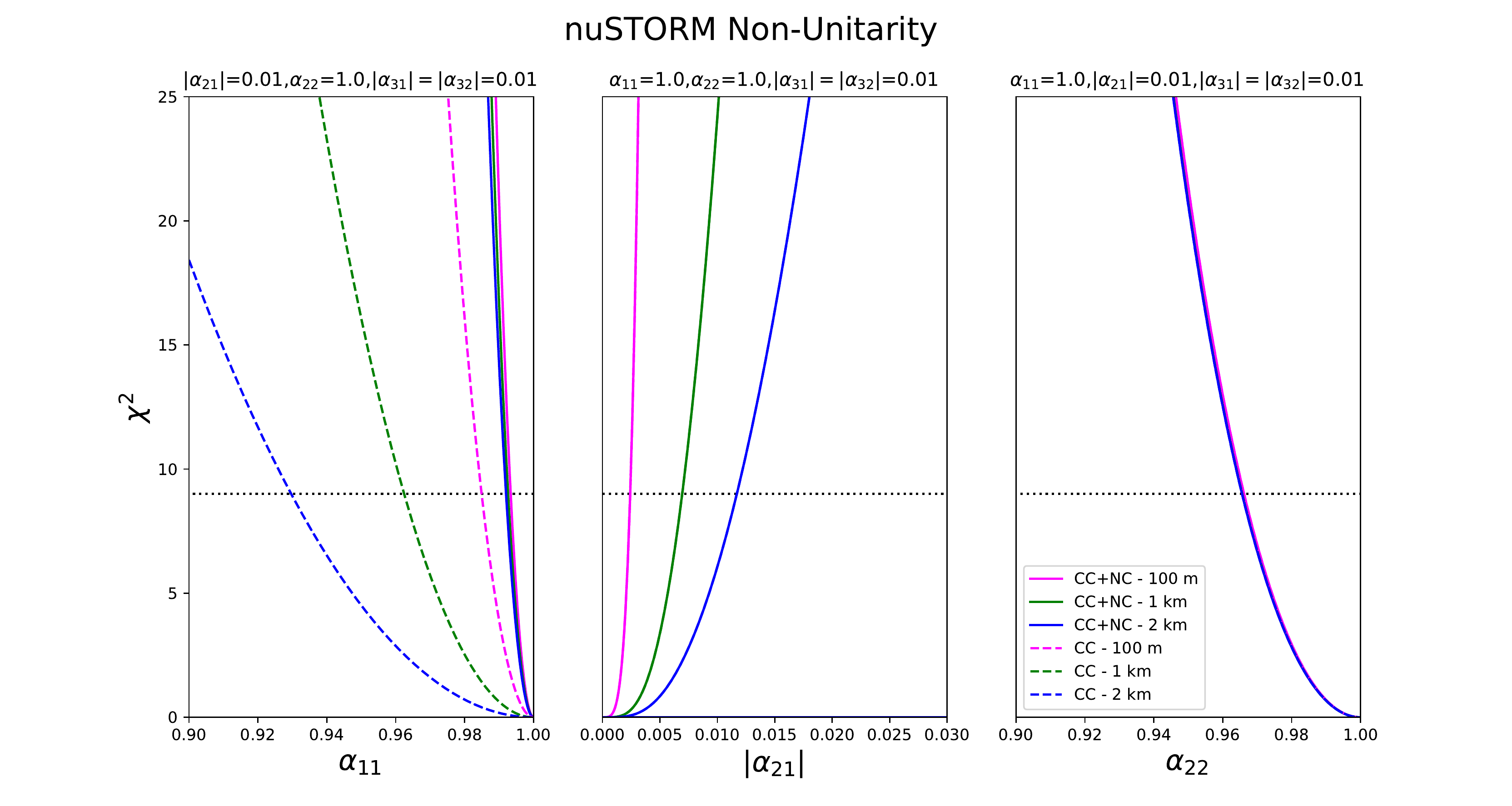} 
\caption{The figure shows the sensitivity to nuSTORM for the non unitarity parameters $\alpha_{11}$, $|\alpha_{21}|$ and $\alpha_{22}$. The y-axis in the plots represent $\chi^2$, while the x-axis denotes $\alpha_{11}$, $|\alpha_{21}|$ and $\alpha_{22}$ for plots respectively. In each plot the  dashed lines are for the contribution of only charge current interactions while the solid lines are for the combination of charge current and neutral current.
The magenta, green and blue curves represent the sensitivities at the baseline of 100 m, 1 km and 2 km respectively.}
\label{fig:nu-par-sens}
%\hspace*{10cm}
\end{figure}
\end{center}

%\begin{center}
%\begin{figure}[h!]
%\hspace*{-2cm}
%\includegraphics[height=6cm]{nustorm-nu-apl11sq-chisq-cc+nc-baseline.eps} 
%\includegraphics[height=6cm]{a11-D1.png} \\
%\includegraphics[height=6cm]{nustorm-nu-apl21sq-chisq-cc+nc-baseline.eps}
%\includegraphics[height=6cm]{a21-D1.png} \\
%\includegraphics[height=6cm]{nustorm-nu-apl22sq-chisq-cc+nc-baseline.eps}
%\includegraphics[height=6cm]{a22-D1.png} \\
%\caption{The figure denotes the sensitivity of NuStorm for the non unitarity parameters $\alpha_{21}$ and $\alpha_{22}$. The y-axis in both plots represent $\chi^2$, while the x-axis denotes $\alpha_{21}$ and $\alpha_{22}$ for left and right panels respectively. In each plot the magenta dashed lines are for the contribution of charge current interactions and the blue solid lines are for the combination of charge current and neutral current.}
%\label{fig:nu-par-sens}
%%\hspace*{10cm}
%\end{figure}
%\end{center}

The capability of nuSTORM to probe the non unitarity parameters $\alpha_{11}$, $|\alpha_{21}|$ and $\alpha_{22}$ are shown in Fig.\ref{fig:nu-par-sens}. Each plot presents 3 cases for 3 different baselines: 100 m; 1 km; and 2 km, plotted with magenta, green and blue curves respectively. The first plot in fig.\ref{fig:nu-par-sens} presents the sensitivity of nuSTORM to the parameter $\alpha_{11}$, with the diagonal parameters $\alpha_{22}=\alpha_{33}=1.0$ and the off-diagonal parameters $|\alpha_{21}|$,$|\alpha_{31}|$ and $|\alpha_{32}|$ fixed at 0.01.%\footnote{These parameters should be kept zero for unitarity, but it has been kept non-zero because taking it zero will provide null contribution from the $P_{e \mu}$ channel.}. 
Unitarity requires that the parameters be set to zero. However, the value of -.01 has been drawn so that a contribution from the $\nu_e \rightarrow \nu_{\mu}$ channel remains. 
Beginning with the first case, which shows the $\chi^2$ as a function of $\alpha_{11}$, under the condition that the parameters $|\alpha_{21}| = 0.1$ and $\alpha_{22} = 1.0$. The true data have been generated keeping $\alpha_{11}$ fixed at unity while the test data have been generated by varying $\alpha_{11}$ between 0.9 and 1.0 while keeping all other parameters fixed.
The relevant channel to study the $\alpha_{11}$ sensitivity is the $P_{e \mu}$ appearance channel because the probability $P_{\bar{\mu} \bar{\mu}}$ is independent of $\alpha_{11}$. Under the above conditions $P_{e \mu} ~ \sim ~ 0.01 \alpha_{11}^2$, therefore, the sensitivity plot has a quadratic dependence on $\alpha_{11}$. From the expression it is clear that $P_{e \mu}$ is independent of the baseline so a change in sensitivity to $\alpha_{11}$ by varying the baseline is due to the change in the flux which occurs due to the change in the baseline. Hence, we observe that the sensitivity increases as the baseline is reduced. $3\sigma$ sensitivity for $\alpha_{11}$ is achieved for $\alpha_{11} = 0.93$ for a 2 km baseline, which increases to $0.96$ for the 1 km baseline and the best result is achieved for the 100 m baseline where the same sensitivity is achieved for $\alpha_{11} = 0.99$. If neutral current events are also combined the charged current events a substantial improvement in the sensitivity is observed. $3\sigma$ sensitivity for $\alpha_{11} = 0.995$ when CC and NC both are taken into consideration.
Similar studies have been performed at DUNE and T2HK \cite{Dutta:2019hmb} where the $3\sigma$ sensitivity for $\alpha_{11} \approx 0.94$ for DUNE and $\alpha_{11} \approx 0.96$ for T2HK. Therefore, we can see that nuSTORM with 2 km baseline has sensitivity similar to DUNE and with baseline 1 km has similar sensitivity to T2HK when the baseline is decreased further the sensitivity increases further exceeding the sensitivities attained by DUNE or T2HK.

The second plot in the Fig.\ref{fig:nu-par-sens} shows the $\chi^2$ vs $|\alpha_{21}|$ sensitivity with both the non-unitarity parameters $\alpha_{11}$ and $\alpha_{22}$ taken to be unity. Under such conditions $P_{e \mu}$ just reduces to $|\alpha_{21}|^2$ and $P_{\bar{\mu} \bar{\mu}}$ becomes $(1+|\alpha_{21}|^2)^2$ which can be approximated to be $\sim~1+2|\alpha_{21}|^2$. Unlike the case discussed above, where only the appearance channel contributes, both the appearance and the disappearance channel contribute to the sensitivity to $|\alpha_{21}|$. Since both the channels depend on $|\alpha_{21}|^2$ we get a quadratic dependence of the $\chi^2$ on $|\alpha_{21}|$. In this case the true data have been generated at $\alpha_{11} = \alpha_{22}= \alpha_{33} =1.0$, $|\alpha_{21}|=|\alpha_{31}|=|\alpha_{32}| = 0$ and $\alpha_{22} = 1.0$, the test data have been generated with $|\alpha_{21}|$ varying in the range 0.0 to 0.01. In this case also we find that the sensitivity is dependent on the baseline for the same reason as discussed previously. The $|\alpha_{21}|$ sensitivity reaches 3$\sigma$ for $|\alpha_{21}| = 0.011$ at the 2 km baseline, $|\alpha_{21}| = 0.006$ at the baseline 1 km, and $|\alpha_{21}| = 0.003$ at the baseline 100 m. Neutral current events do not contribute to the $|\alpha_{21}|$ sensitivity, this is because $P_{NC}~\approx~ 2 - (\alpha_{11}^2 + \alpha_{22}^2)(\alpha_{11}^2 + 2|\alpha_{21}|^2 +|\alpha_{31}|^2 )$ where $(\alpha_{11}^2 + 2|\alpha_{21}|^2 +|\alpha_{31}|^2 ) \approx 1$ as a result the NC channel cannot probe $|\alpha_{21}|$ independently. Comparing the sensitivities with DUNE and T2HK \cite{Dutta:2019hmb} we observe that nuSTORM can reach $3\sigma$ sensitivity for $|\alpha_{21}|$ for an order of magnitude smaller values of $|\alpha_{21}|$. nuSTORM has a significant advantage over DUNE and T2HK which can reach $3\sigma$ sensitivities for $|\alpha_{21}| = 0.08$ and $0.04$ respectively.

The third figure presents the sensitivity to the parameter $\alpha_{22}$. $P_{e \mu}$ is independent of $\alpha_{22}$ but $P_{\bar{\mu} \bar{\mu}}$ is sensitive to $\alpha_{22}$. The true data has been generated by considering unitary evolution i.e. $\alpha_{11} = 1.0$, $|\alpha_{21}| = 0$ and $\alpha_{22} = 1.0$ which reduces $P_{\bar{\mu} \bar{\mu}}$ to $\alpha_{22}^4$. The test data have been generated by taking $\alpha_{11} = 1.0$, $|\alpha_{21}| = 0$ and varying $\alpha_{22}$ from 0.9 to 1.0. An interesting feature observed here is the independence of $\alpha_{22}$ on the baseline. This can be attributed to the fact that the sensitivity is solely dependent on the disappearance channel which already has enough statistics at 2 km, hence reducing the baseline does not help. 
The introduction of neutral current events is expected to increase the $\alpha_{22}$ sensitivity because of the dependence of $P_{NC}$ on $\alpha_{22}^2$. However, no improvement is observed because the introduction of the channel increases the statistics but it already had enough statistics from the disappearance channel itself.
The 3$\sigma$ sensitivity is reached at $\alpha_{22} = 0.97$ for all baselines. Again from \cite{Dutta:2019hmb}, the $3\sigma$ sensitivities for DUNE and T2HK for $\alpha_{22}$ can be attained for $\alpha_{22} \approx 0.98$ for both the experiments.

\section{Conclusions} \label{conc}
In this work we have investigated the capabilities of nuSTORM to explore two new physics scenarios -- (i) the existence 
of $\rm{eV^2}$ scale oscillation, suggested as an explanation of 
LSND/MiniBOONE anomalies and (ii) non-unitarity of the neutrino 
mixing matrix.   
nuSTORM is  proposed primarily to measure the $\nu_e ~ N$ and $\nu_\mu ~ N$ cross sections. 
It was shown in \cite{Tunnell:2012nu, Adey:2014rfv} that nuSTORM can also play in important role to study active-sterile oscillations governed by an $\rm{eV^2}$ mass 
squared difference.  In this work, we have studied  the effect of 
including neutral current events and checked whether this can give improved 
sensitivity to sterile-neutrino searches.
nuSTORM will have the capability to study two main channels,
the conversion probability  $P_{\mu e}$ and survival probability 
$P_{\bar{\mu} \bar{\mu}}$ with the proposed MIND detector. 
Whereas, for oscillations involving active neutrinos the NC events 
are not sensitive to oscillation parameters, for oscillations involving the sterile neutrinos, the neutral current events are also 
sensitive to the oscillation parameters through the probabilities
involving conversion to sterile neutrinos $P_{\mu s}$ and $P_{e s}$.  
Considering a 2 km baseline it is observed that taking only
CC interactions can constrain the mixing angle 
$\theta_{24} \lesssim 7.5^\circ$ but cannot constrain $\theta_{24}$,  which can be achieved with the inclusion of NC interactions. 
For non-zero values of $\theta_{24}$, 
the constraint on $\theta_{14}$ also improves with inclusion 
of NC events.  
Since, nuSTORM is a proposed experiment, baseline optimization is important to maximize physics output. When we consider various baselines we find that the baseline of 1 km gives a good overall sensitivity for both $\theta_{14}$ and $\theta_{24}$ over a wide range of $\Delta m^2_{41}$.

For the other new-physics scenario, non-unitarity of the lepton mixing matrix, studied in this work, we find that
nuSTORM can probe the non-unitarity parameters $\alpha_{11}$, $|\alpha_{21}|$ and $\alpha_{22}$. 
$3\sigma$ sensitivities for $\alpha_{11}$, $|\alpha_{21}|$ and $\alpha_{22}$ are obtained at 0.995, 0.011 and 0.97 respectively for 2 km baselines combining both CC and NC events. The sensitivities for $\alpha_{11}$ and $|\alpha_{21}|$ significantly improves as the baseline is reduced. 

In conclusion, we find that apart from measuring neutrino cross-sections with per mil precision, nuSTORM can also contribute significantly by probing new physics scenarios beyond Standard Model . 

\section*{Acknowledgements}

S. Goswami acknowledges Leverhulme Trust visiting Professorship and the hospitality at Imperial  College London where the work was done. K. Chakraborty thanks the support provided by the Indo-French Centre for the Promotion of Advanced Research (IFCPAR/CEFIPRA) for the Raman-Charpak Fellowship 2018 and the Institut Pluridisciplinaire Hubert Curien, IPHC, Strasbourg for their kind hospitality where a part of the work was completed.

%%%%%%%%%%%%%%%%%   bibliography %%%%%%%%%%%%
\bibliographystyle{utphys}
\bibliography{sterile-ref}

\end{document}